%% file: hadatom05_proc.tex
\documentclass[12pt]{article}
\usepackage{a4}
\usepackage{latexsym} 
\usepackage{amssymb}  
\usepackage{amsmath} 
\usepackage{graphicx}
\usepackage{epsfig,wrapfig}
\usepackage[ usenames]{color}
\usepackage[rflt]{floatflt} 
\begin{document}

\newcommand{\talk}[3]
{\noindent{#1}\\ \mbox{}\ \ \ {\it #2} \dotfill {\pageref{#3}}\\[1.8mm]}
\newcommand{\stalk}[3]
{{#1} & {\it #2} & {\pageref{#3}}\\}
\newcommand{\snotalk}[3]
{{#1} & {\it #2} & {{#3}n.r.}\\}
\newcommand{\notalk}[3]
{\noindent{#1}\\ \mbox{}\ \ \ {\it #2} \hfill {{#3}n.r.}\\[1.8mm]}
\newcounter{zyxabstract}     
\newcounter{zyxrefers}        

\newcommand{\newabstract}
{\clearpage\stepcounter{zyxabstract}\setcounter{equation}{0}
\setcounter{footnote}{0}\setcounter{figure}{0}\setcounter{table}{0}}

\newcommand{\talkin}[2]{\newabstract\label{#2}\input{#1}}                

\newcommand{\rlabel}[1]{\label{zyx\arabic{zyxabstract}#1}}
\newcommand{\rref}[1]{\ref{zyx\arabic{zyxabstract}#1}}

\renewenvironment{thebibliography}[1] 
{\section*{References}\setcounter{zyxrefers}{0}
\begin{list}{ [\arabic{zyxrefers}]}{\usecounter{zyxrefers}}}
{\end{list}}
\newenvironment{thebibliographynotitle}[1] 
{\setcounter{zyxrefers}{0}
\begin{list}{ [\arabic{zyxrefers}]}
{\usecounter{zyxrefers}\setlength{\itemsep}{-2mm}}}
{\end{list}}

\renewcommand{\bibitem}[1]{\item\rlabel{y#1}}
\renewcommand{\cite}[1]{[\rref{y#1}]}      
\newcommand{\citetwo}[2]{[\rref{y#1},\rref{y#2}]}
\newcommand{\citethree}[3]{[\rref{y#1},\rref{y#2},\rref{y#3}]}
\newcommand{\citefour}[4]{[\rref{y#1},\rref{y#2},\rref{y#3},\rref{y#4}]}
\newcommand{\citefive}[5]
{[\rref{y#1},\rref{y#2},\rref{y#3},\rref{y#4},\rref{y#5}]}
\newcommand{\citesix}[6]
{[\rref{y#1},\rref{y#2},\rref{y#3},\rref{y#4},\rref{y#5},\rref{y#6}]}

\begin{titlepage}

\begin{flushleft}
\includegraphics[height=1.5cm]{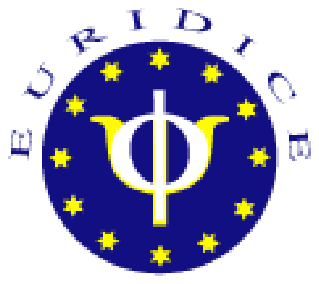} 
\end{flushleft}

\vspace*{-1.75cm}
\begin{flushright}
\small{\tt Bern University, CERN and JINR Dubna}
\end{flushright}

\vspace*{1.cm}

\begin{center}
  {\Huge \bf HadAtom05}\\[0.5cm]
  {\large\bf Workshop on Hadronic Atoms},\\
  University of Bern, Switzerland\\
  February 15 and 16, 2005\\[1cm]

\begin{center}
\includegraphics[height=5.cm]{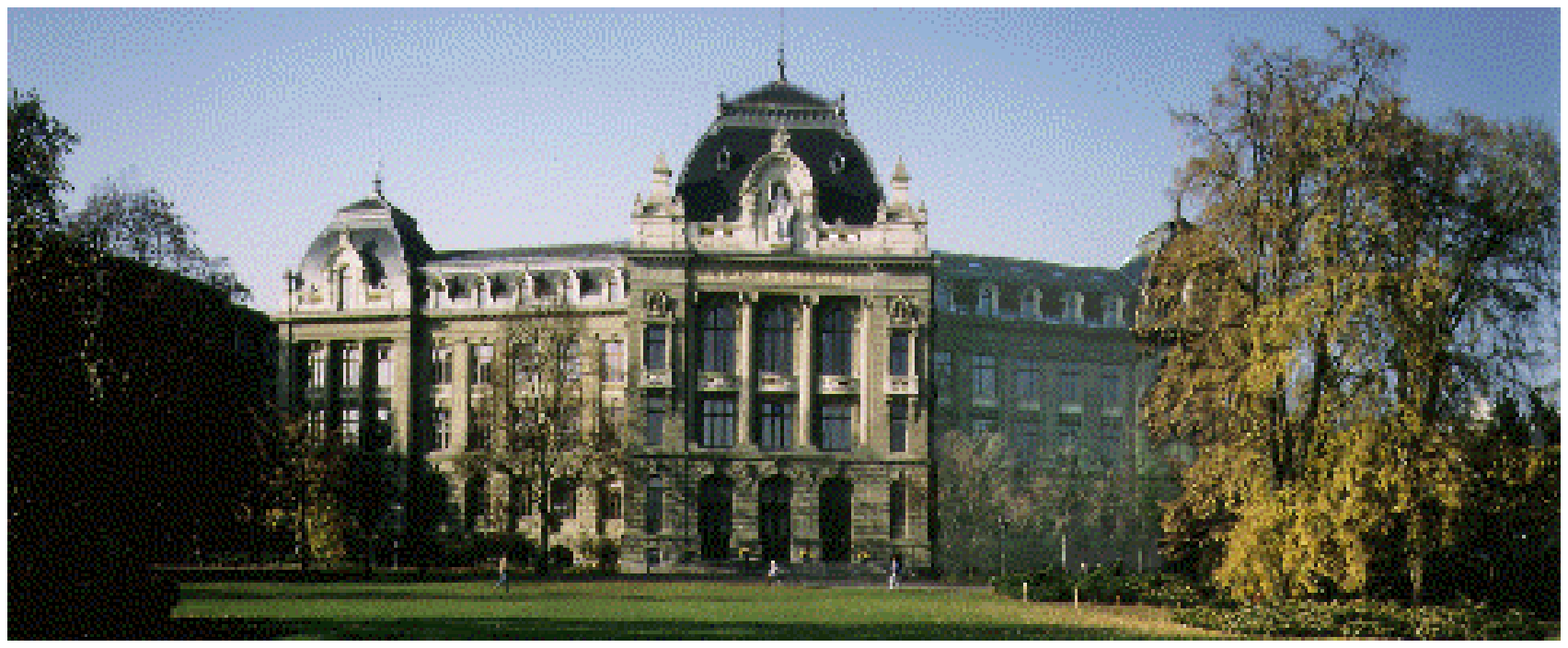} 
\end{center}

  {\em edited by}\\[1cm]
  {\bf L. Afanasyev$^1$, G. Colangelo$^2$ and J. Schacher$^3$}\\[0.3cm]
  {\em $^1$Joint Institute for Nuclear Research, 141980 Dubna,Moscow Region, Russia}\\
  {\em $^2$ Institute for Theoretical Physics, University of Bern,
    Sidlerstrasse 5, 3012 Bern, Switzerland}\\
  {\em $^3$Laboratory for High-Energy Physics, University of Bern,
    Sidlerstrasse 5, 3012 Bern, Switzerland}

\end{center}

\vspace*{1.0cm}

\begin{abstract}
  \baselineskip 1.5em These are the proceedings of the workshop
  ``HadAtom05'', held at the University of Bern, Switzerland, February
  15 and 16, 2005.  The main topics of the workshop concerned the
  physics of hadronic atoms, and in this context recent results from
  experiments and theory were presented.  These proceedings contain
  the list of participants, the scientific program and a short
  contribution from each speaker.
\end{abstract}

\end{titlepage}

\setcounter{page}{2}

\newabstract

\section{Introduction}

The workshop ``HadAtom05'' was held at the University of Bern, 
Switzerland, on February 15 and 16, 2005.  This was already the 
sixth regular workshop of the HadAtom type. This workshop series 
has originally been inspired by the latest experimental and 
theoretical progress achieved in the investigation of the bound 
states of strongly interacting particles, the hadronic atoms. 
The previous workshops were held in Dubna (1998), Bern
(1999, 2001), at CERN (2002) and in Trento (2003). 
The present workshop covered the following topics:

\begin{itemize}
\itemsep -1.mm
\item
Hadronic atoms, in particular their
\vspace{-1ex}
\begin{itemize}
\itemsep -1.5mm
 \item               Production
 \item               Interaction with matter
 \item               Energy levels
 \item               Decays
\end{itemize}
\vspace{-1ex}

\item
Meson-meson and meson-baryon scattering

\item
Lattice calculations and low energy effective theory of QCD

\item
Chiral Perturbation Theory and nuclear many-body systems
\vspace{-1ex}
\begin{itemize}
\itemsep -1.5mm
\item                Nuclear Matter
\item                Pionic deuteriun and pionic hydrogen
\end{itemize}
\vspace{-1ex}

\item
Experiments
\vspace{-1ex}
\begin{itemize}
\itemsep -1.5mm
\item                DIRAC at CERN
\item                DEAR/SIDDHARTA at DAFNE
\item                Pionic Hydrogen Collaboration at PSI
\item                Hadronic atoms at J-PARC and GSI
\item                $K$ decays with NA48 at CERN
\end{itemize}
\vspace{-1ex}

\end{itemize}

About 40 physicists took part in the workshop, and 23 talks were
presented.  As for the previous workshops
\citefive{98}{99}{01}{02}{03}, we publish a collection of abstracts of
the presentations, containing relevant references. In addition, we
display the list of the participants with their e-mail addresses.

\bigskip\bigskip

{\em Acknowledgments}

We would like to thank all participants for their effort 
to travel to Bern and for making ``HadAtom05'' an exciting 
meeting. We cordially thank our secretaries, Ruth Bestgen 
and Ottilia H\"anni, for the excellent performance, and
the staff of the Institut f\"ur Theoretische Physik for 
their invaluable support. Last but not least, we also thank
our colleagues from the program panel, namely J\"urg Gasser, 
Heinrich Leutwyler, Leonid Nemenov, Akaki Rusetsky and 
Dirk Trautmann, for their continuous help in structuring 
this workshop.

\bigskip

\noindent Bern and CERN, July 2005

\bigskip\bigskip\bigskip

\hspace*{10.cm} Leonid Afanasyev

\hspace*{10.cm} Gilberto Colangelo

\hspace*{10.cm} J\"urg Schacher

\newpage
\section{List of Particpants}
\begin{tabular}{rlll}
 1 & Afanasyev Leonid & Dubna & afanasev@nu.jinr.ru\\
 2 & Allkofer Yves & Zurich & Yves.Allkofer@cern.ch\\
 3 & Aste Andreas & Basel & aste@quasar.physik.unibas.ch\\
 4 & Cabibbo Nicola & Roma & nicola.cabibbo@roma1.infn.it\\
 5 & Colangelo Gilberto & Bern & gilberto@itp.unibe.ch\\
 6 & dos Santos Covita Daniel  & PSI & daniel.covita@psi.ch\\
 7 & Dreyer Ute & Basel & dreyer@quasar.physik.unibas.ch\\
 8 & Friedman  Eli &  Jerusalem  & elifried@vms.huji.ac.il\\
 9 & Gasser Juerg & Bern & gasser@itp.unibe.ch\\
10 & Goldin Daniel & Basel & daniel.goldin@cern.ch\\
11 & Gotta Detlev & Jülich & d.gotta@fz-juelich.de\\
12 & Heim Thomas & Basel  & thomas.heim@unibas.ch\\
13 & Hencken Kai & Basel & k.hencken@unibas.ch\\
14 & Ivanov Andrei & Vienna  & ivanov@kph.tuwien.ac.at\\
15 & Jensen Thomas  & Paris & Thomas.Jensen@spectro.jussieu.fr\\
16 & Johnson Ian & Zurich & Ian.Johnson@cern.ch\\
17 & Juge Jimmy & Dublin & juge@maths.tcd.iet\\
18 & Kubis Bastian & Bonn  & kubis@itp.unibe.ch\\
19 & Lemmer Richard & Johannesburg  & iclemmer@iafrica.com\\
20 & Lyubovitskij Valery & Tuebingen  & valeri.lyubovitskij@uni-tuebingen.de\\
21 & Mannelli Italo & PISA & mannelli@pi.infn.it\\
22 & Marton Johann & Vienna & johann.marton@oeaw.ac.at\\
23 & Minkowski & ITP, Bern & mink@itp.unibe.ch\\
24 & Nemenov Leonid & CERN--JINR & Leonid.Nemenov@cern.ch\\
25 & Pineda Antonio & Barcelona &  pineda@ecm.ub.es\\
26 & Raha Udit & Bonn & udit@itkp.uni-bonn.de\\
27 & Rusetsky Akaki & Bonn  & rusetsky@itkp.uni-bonn.de\\
28 & Sainio Mikko & Helsinki  & mikko.sainio@helsinki.fi\\
29 & Santos, José Paulo & Lisboa   & jps@cii.fc.ul.pt\\
30 & Sazdjian Hagop & Paris  & sazdjian@ipno.in2p3.fr\\
31 & Schacher Juerg & Bern  & juerg.schacher@lhep.unibe.ch\\
32 & Simons   Leopold & PSI & leopold.simons@psi.ch\\
33 & Tarasov Alexandr &  Dubna & tarasov@nu.jinr.ru\\
34 & Torleif Ericson & CERN & torleif.ericson@cern.ch\\
35 & Trautmann Dirk & Basel & Dirk.Trautmann@unibas.ch\\
36 & Voskresenskaya Olga & Dubna & Olga.Voskresenskaja@mpi-hd.mpg.de\\
37 & Yaz'kov Valeriy & Dubna & Valeri.Iazkov@cern.ch\\
38 & Zemp Peter & Bern  & zemp@itp.unibe.ch\\
39 & Zhabitsky Mikhail & Dubna & zhabitsk@nusun.jinr.ru\\
\end{tabular}

\newpage

\section{Scientific program}

\vskip.5cm

\noindent\mbox{}\hfill{\bf Page}

\talk{{\bf M.~Zhabitsky}}{The DIRAC experiment at CERN}{abs:zhabitsky}
\talk{{\bf L. Nemenov}}{Using $\pi^+\pi^-$ and $\pi K$ atoms for the experimental check of the low-energy QCD}{abs:nemenov}
\talk{{\bf L. Afanasyev} and L. Nemenov}
{Production of the long-lived excited states of $\pi^+\pi^-$ atoms for DIRAC}{abs:afanasev}
\talk{T. Heim, {\bf K. Hencken}, M. Schumann, D. Trautmann and G. Baur}
{Electromagnetic Exctiation and Ionization of Pionium: An Update}{abs:hencken}
\talk{{\bf J.~Juge}}{$\pi\pi$ threshold parameters from lattice calculations}{abs:juge}
\talk{{\bf D. Gotta}}{The pionic hydrogen experiment at PSI}{abs:gotta}
\talk{Ulf-G. Mei{\ss}ner, Udit Raha and {\bf Akaki Rusetsky}}
{The $\pi N$ scattering lengths from pionic deuterium}{abs:rusetsky}
\talk{{\bf M.E. Sainio}}{Progress report in $\pi$N analysis}{abs:sainio}
\talk{{\bf P. Zemp}}{Decay width of the ground state of pionic hydrogen in QCD+QED}{abs:zemp}
\talk{{\bf T. E. O. Ericson} and A. N. Ivanov}
{The $\pi ^-p$ atomic level shift and  Isospin Violation in $\pi N$ Scattering}{abs:ericson}
\talk{{\bf T.S. Jensen}}{Kinetic energy distributions in pionic hydrogen}{abs:jensen}
\talk{{\bf A. Pineda}}{The chiral structure of the (muonic) hydrogen}{abs:pineda}
\talk{{\bf J. Marton}}{Precision Measurements with Kaonic Atoms}{abs:marton}
\talk{{\bf U. Raha}}{Kaonic hydrogen}{abs:raha}
\notalk{{\bf J. Santos}} {Kaonic and sigmonic atoms}{~}
\talk{{\bf A.~N.~Ivanov},M.~Cargnelli, M.~Faber,~H.~Fuhrmann, 
 V.~A.~Ivanova, J.~Marton, N.~I.~Troitskaya, and J.~Zmeskal}{On the energy level displacement of the
    ground state of kaonic deuterium}{abs:ivanov}
\talk{Amand\ Faessler, Th.\ Gutsche, {\bf V.\ E.\ Lyubovitskij}, K. \ Pumsa-ard}
{Nucleon stucture in Lorentz covariant chiral quark model}{abs:lyubovitskij}
\talk{{\bf I. Mannelli}}{Study of pion-pion interaction in $K \to 3 \pi$ decays}{abs:mannelli}
\notalk{{\bf N. Cabibbo}}{Cusp effects in $K\to3\pi$ and
the determination of $\pi\pi$ scattering lengths}{~}
\talk{S. Bakmaev, {\bf A. Tarasov}, and O. Voskresenskaya}
{The unitary correction to the Moli\`ere--Fano multiple scattering theory}{abs:tarasov}
\talk{S. Bakmaev  and O. Voskresenskaya}
{Production of pioniums in the coherent nuclear-nuclear collisions}{abs:voskr}
\talk{{\bf E. Friedman}}{Pion-nucleus interaction at low energy}{abs:friedman}
\talk{{\bf J. Gasser}}{Concluding remarks}{abs:gasser}
\hspace*{1.cm}


\newabstract\label{abs:zhabitsky}\input{zhabitsky}

\newabstract\label{abs:nemenov}\input{nemenov}

\newabstract\label{abs:afanasev}\input{afanasev}

\newabstract\label{abs:hencken}\input{hencken}

\newabstract\label{abs:juge}\input{juge}

\newabstract\label{abs:gotta}\input{gotta}

\newabstract\label{abs:rusetsky}\input{rusetsky}

\newabstract\label{abs:sainio}\input{sainio}

\newabstract\label{abs:zemp}\input{zemp}

\newabstract\label{abs:ericson}\input{ericson}

\newabstract\label{abs:jensen}\input{jensen}

\newabstract\label{abs:pineda}\input{pineda}

\newabstract\label{abs:marton}\input{marton}

\newabstract\label{abs:raha}\input{raha}

\newabstract\label{abs:ivanov}\input{ivanov}

\newabstract\label{abs:lyubovitskij}\input{lyubovitskij}

\newabstract\label{abs:mannelli}\input{mannelli}

\newabstract\label{abs:tarasov}\input{tarasov}

\newabstract\label{abs:voskr}\input{voskr}

\newabstract\label{abs:friedman}\input{friedman}

\newabstract\label{abs:gasser}\input{gasser}

\end{document}

%% file: zhabitsky.tex
\begin{center}
{\large\bf The DIRAC experiment at CERN}\\[0.5cm]
M.~Zhabitsky on behalf of DIRAC collaboration\\[0.3cm]
Joint Institute for Nuclear Research, Joliot-Curie 6, 141980 Dubna, Russia
\end{center}

The aim of the DIRAC experiment at CERN~\cite{DIRAC_proposal:1994}
is to measure the lifetime of pionium, a hydrogen-like atom consisting 
of a $\pi^+$ and a $\pi^-$ meson.
The lifetime $\tau$ is dominated by the charge exchange process 
($\pi^+\pi^-\rightarrow\pi^0\pi^0$) and thus
is inversely proportional to the squared difference
between S-wave $\pi\pi$~scattering lengths with isospin 0 and 2:
$|a_0-a_2|$.
This difference was calculated within the framework
of Standard Chiral Perturbation Theory with a precision
$1.5$\% ~\cite{ChPT_Colangelo:2001}
which leads to the lifetime prediction
$\tau_{1S}=(2.9\pm 0.1)\times 10^{-15}$s.

The DIRAC exploits the lifetime measurement method proposed
in~\cite{Nemenov:1985}. 
Pionium atoms are produced in proton-nucleus interactions. 
After production these relativistic atoms may either decay into 
$\pi^0\pi^0$ or get excited to higher quantum numbers, or break-up
in the target material. 
In the case of break-up, characteristic pion pairs $n_A$ emerge
which have a low relative momentum~$Q$ in their centre of mass system.
The number of produced atoms $N_A$ is proportional
to the number of free pion pairs with small $Q$ which undergo
Coulomb interaction in final state ${N_{CC}}$.
Then for a given target material 
there is a one-to-one correspondence between the measured
break-up probability $P_\text{br}$ and 
atomic lifetime~\cite{atoms_interact}.

In 2001 DIRAC collaboration obtained
a first high statistics ($\sim\! 10^4$) atomic data 
sample collected from p-Ni interactions 
at $24\:$GeV/c~\cite{DIRAC_observation:2004}.
The analysis of experimental distributions on $\pi\pi$~pairs relative momentum
results in $n_A^{rec}=6530\pm 294$ pairs from pionium break-up.
Then the break-up probability $P_\text{br}$ 
becomes [to be published]:
\begin{equation}
 P_\text{br} = \frac{n_A}{N_A} 
             = \frac{n^{rec}_A(Q\leqslant Q_\text{cut})}
                    {k(Q_\text{cut})N^{rec}_{CC}(Q\leqslant Q_\text{cut})}
             = 0.452\pm0.023|_\text{stat},
\end{equation}
which corresponds to the lifetime of the atomic ground state
\begin{equation}
\tau_{1S}= \left.\left.2.91^{+0.45}_{-0.38}\right|_\text{stat}
                         {}^{+0.19}_{-0.49}\right|_\text{syst}\times 10^{-15}s.
\end{equation}

The systematical error is dominated by uncertainties in multiple scattering,
presence of non-identified admixture of $K^+K^-$ and proton-antiproton pairs,
uncertainties in the correction due to finite size production and
strong interaction for Coulomb-correlated pairs.
The analysis of the full data sample (about 2 times more than analysed here) 
and dedicated measurements, 
which will decrease systematical errors, 
is in progress.

%% file: nemenov.tex
\begin{center}
{\large\bf Using $\pi^+\pi^-$ and $\pi K$ atoms for the experimental check of the low-energy QCD}\\[0.5cm]
  {\bf L.Nemenov}$^{1,2}$\\[0.3cm]
  $^1$CERN, CH-1211 Geneva 23, Switzerland\\[0.1cm]
  $^2$Joint Institute for Nuclear Research, 141980 Dubna,Moscow Region, Russia\\[0.1cm]
\end{center}

The lifetime ($\tau$) measurement of atoms consisting of $\pi^+\pi^-$
($A_{2\pi}$) and $\pi K$ mesons ($A_{\pi K}$) allows to obtain in a
model independent way the difference of $\pi\pi$ $(a_0-a_2)$ and $\pi
K$ $(a_{1/2}-a_{3/2})$ scattering lengths in S-state.  Basing on the
last experimental data on $\tau(A_{2\pi})$ from DIRAC experiment at
CERN the expected precisions of this parameter for the approved
upgrade of the experiment are presented in Tables~\ref{nem:tab1} and
\ref{nem:tab2}.  The expected accuracy of $\tau(A_{2\pi})$ and
$\tau(A_{\pi K})$ measurement on J-PARC, GSI and SPS CERN are
presented in the same tables also.  Observation of the long-lived
(metastable) states of $A_{2\pi}$ opens the possibility to measure the
Lamb-shift in $A_{2\pi}$ and the measurement the other combination the
$\pi\pi$ scattering lengths.

\begin{table}[h]
  \caption{Statistics of $A_{2\pi}$ ($n_A$) and the corresponding
    statistical error (stat) in $|a_0-a_2|$ to be obtained with the
    upgraded DIRAC setup during 12 month runs (20h/day) are shown for
    different accelerators. The latter value should be compared to the
    contemporary limits in accuracy of $|a_0-a_2|$ coming from the
    utilized theoretical dependences of $\tau(A_{2\pi})$ on the
    $|a_0-a_2|$ ($\tau=f(a_0-a_2)$) and the probability of the $A_{2\pi}$
    breakup in the target on $\tau$ ($P_{br}=f(\tau)$).}
  \label{nem:tab1}
  \centering
  \begin{tabular}{|l|cccc|}
\hline
& $n_A$ & stat & $\tau=f(a_0-a_2)$ & $P_{br}=f(\tau)$\\
\hline
PS CERN & 85000 & 2\% & 0.6\% & 1.2\% \\
24 Gev/$c$ & & & &\\
J-PARC  & $4.1\times10^5$ & 0.9\% & 0.6\% & 1.2\% \\
50 Gev/$c$ & & & &\\
GSI & $1.2\times10^6$ & 0.6\% & 0.6\% & 1.2\% \\
90 Gev/$c$ & & & &\\
SPS CERN & $1.26\times10^6$ & 0.5\% & 0.6\% & 1.2\% \\
24 Gev/$c$ & & & &\\
\hline
  \end{tabular}
\end{table}

\begin{table}[h]
  \caption{The same values as in Table~1 for $A_{\pi K}$ atoms and
    $|a_{1/2}-a_{3/2}|$ value.}
  \label{nem:tab2}
  \centering
  \begin{tabular}{|l|cccc|}
\hline
& $n_A$ & stat & $\tau=f(a_{1/2}-a_{3/2})$ & $P_{br}=f(\tau)$\\
\hline
PS CERN & 7000 & 10\% & 1.1\% & 1.2\% \\
24 Gev/$c$ & & & &\\
J-PARC  & $4.1\times10^5$ & 7\% & 1.1\% & 1.2\% \\
50 Gev/$c$ & & & &\\
GSI & $1.2\times10^6$ & 2.5\% & 1.1\% & 1.2\% \\
90 Gev/$c$ & & & &\\
SPS CERN & $1.26\times10^6$ & 2.5\% & 1.1\% & 1.2\% \\
24 Gev/$c$ & & & &\\
\hline
  \end{tabular}
\end{table}

%% file: afanasev.tex
\begin{center}
  {\large\bf Production of the long-lived excited states of $\pi^+\pi^-$
    atoms for DIRAC}\\[0.5cm]
{\bf L. Afanasyev} and L. Nemenov\\[0.3cm]
Joint Institute for Nuclear Research, 141980 Dubna,Moscow Region, Russia
\end{center}

Measurement of the $\pi^+\pi^-$ atom lifetime in the DIRAC experiment
will allows one to obtain in the model independent way the value
$|a_0-a_2|$, the difference of the $s$-wave $\pi\pi$ scattering
lengths with the isotop spin 0 and 2 correspondently \cite{dirac}. To
get the values of $a_0$ and $a_2$ separately basing on the
$\pi^+\pi^-$ atom data, one may use the fact that the energy splitting
between levels $ns$ and $np$, $\Delta E_n=E_{ns}-E_{np}$, depends on the
the another combination of scattering lengths: $2a_0+a_2$. Thus the
measurement the energy splitting coupled with the lifetime measurement
would provide a determination of $a_0$ and $a_2$ separately
\citetwo{dirac}{nem85}.

The lifetimes of the $np$ states are significantly, 3--5 order, higher
in compare with the ground state \cite{nem85}. For that reason atoms
in $np$ states have the mean paths of teens centimeters. Methods of
$\Delta E_n$ measurement proposed in papers \cite{nem01} and
\cite{nem02} based on observation of the interference between $ns$ and
$np(m=0)$ states in the external electro-magnetic fields.

Production of $\pi^+\pi^-$ atom in the $np$ states have been
considered for different target materials and thicknesses with the
intent to optimize the experimental conditions for their
observation. It have been shown that for the DIRAC experiment usage of
thiner targets with smaller $Z$ provides increase of yield of $np$
states and a better ratio to the atom break-up. In the following table
a set of targets providing the highest yield of  $\pi^+\pi^-$ atom
states with the magnetic quantum number $l\ge1$ are shown.

\vspace{1mm}
\noindent \begin{tabular}{|c|c|c|c|c|c|c|c|}
\hline
Target & Thickness & Break-up & $\sum (l\ge1)$ & $2p_0$ & $3p_0$ & $4p_0$ &  $\sum(l=1,m=0)$\\
Z & $\mu m$ &&&&&& \\
\hline
04 & 50 & 2.63\% & 5.86\% & 1.05\% & 0.54\% & 0.20\% & 1.93\% \\
06 & 50 & 5.00\% & 6.92\% & 1.46\% & 0.51\% & 0.16\% & 2.52\% \\
13 & 20 & 5.28\% & 7.84\% & 1.75\% & 0.57\% & 0.18\% & 2.63\% \\
28 & 5  & 9.42\% & 9.69\% & 2.40\% & 0.58\% & 0.18\% & 3.29\% \\
78 & 2  & 18.8\% & 10.5\% & 2.70\% & 0.55\% & 0.16\% & 3.53\% \\
\hline
\end{tabular}
\vspace{1mm}

The first estimation of expected accuracy in the energy splitting
shows that for the magnetic filed of 1~T and 10~cm long measurement
during 6 months (3+3) at PS CERN will provide accuracy in about 6\%
for $\pi^+\pi^-$ atom. For this calculation only atomic pairs was
considered as the background.

%% file: hencken.tex
\begin{center}
{\large\bf Electromagnetic Exctiation and Ionization of Pionium: \\ An Update}\\[0.5cm]
T. Heim$^1$, {\bf K. Hencken}$^1$, M. Schumann$^1$, D. Trautmann$^1$ and G. Baur$^2$\\[0.3cm]
$^1$Institut f\"ur Physik, University of Basel, Klingelbergstr. 82, 4056 Basel, Switzerland\\[0.1cm]
$^2$Institut f\"ur Kernphysik. Forschungszentrum J\"ulich, 52425 J\"ulich, Germany\\[0.1cm]
\end{center}

The precise knowledge of excitation and ionization cross sections of
pionium going through matter is an important theoretical input into
the analysis of the DIRAC experiment \cite{Adeva:2004bd}. In order to achieve
an accuracy of 5\% for the lifetime of the pionium the cross section
needs to be known to a high precision better than 1\%.

At this accuracy a number of additional processes must be taken
into account. This has been achieved by our group in a series of
papers. The dominant contribution coming from the electric interaction
has been studied within a semiclassical treatment in \cite{Halabuka:1999bm}.
There, we calculated total 
and partial cross sections and we studied the influence of
atomic screening. The inclusion of target inelastic
processes, that is ``antiscreening'', was investigated in \cite{Heim00}
and was found to be important (with a correction up to 13\%) 
for light target atoms. The magnetic
interaction, the $A^2$ interaction, and also additional relativistic
corrections were found to be below the required 1\% level
\cite{Heim01}. Multiphoton exchange corrections were calculated within
Glauber theory; they reduce the differential cross section by as much as 30\%
for heavy targets \citetwo{Schumann02}{Schumann03}.

In order to test the accuracy of the Glauber calculations we have
tried to calculate some important cross sections within a coupled
channel approach \cite{Schumann03}. It was found that in a calculation 
taking only the bound states into account, the effect of higher
orders is substantially underestimated. Inclusion of the continuum
states by discretizing them with Weyl wave packets was found to 
give a correction in the right direction, but the overall agreement
with the Glauber calculation is still poor. Therefore, as an
alternative approach we are now studying the possibility of including
the continuum states either within a perturbative treatment, or
within a polarisation
potential model, where the high lying continuum states can be
approximately integrated out. With the help of the closure
approximation only matrix elements between bound states need to be
evaluated. This work is still in progress.

A peculiarity occurs in calculating the deexcitation cross
section, where the pionium undergoes a $2p\rightarrow1s$ transition 
while the atom is excited. Due to the relativistic motion one may encounter 
a singularity of the photon propagator $1/Q^2$.
The integration over $Q$ then leads to a
cross section that is formally divergent. This problem
is not new, of course,  going back to Fano \cite{Fano63}. More
recently it has been studied in connection with a future muon collider,
in the process $\mu^- + \mu^+ \rightarrow e+ \bar\nu + W^+$ through 
a $\nu_\mu$ 
\citetwo{Ginzburg95}{Melnikov96} and in connection with atomic physics
processes \cite{Voitkiv04}. Different scenarios lead to a
regularisation of the photon propagator.

In the case of the DIRAC experiment two phenomena are of
importance, having to do either with the finite lifetime of the
excited pionium beam, or with the finite lifetime of the propagating
photon. The $2p$ state of pionium is not stable but can ``decay''
either by a transition to the $1s$ state, or by interactions with the
target atoms, leading, e.g., to further excited states. Both processes lead
to an
imaginary part of the mass of the $2p$ state which then translates to
an imaginary part for the photon propagator 
\[
q^2 \rightarrow q^2 +
\frac{i}{2} m \Gamma,
\]
with $m$ the mass and $\Gamma$ the width of the
$2p$ states. 
We find that the width due to the
additional interaction is of the order of 1\,eV and therefore the
dominant effect.

For the photon one gets a finite lifetime due to its attenuation in
matter. Starting from the telegraph equation we have derived the
modification to the propagator which again 
acquires an
imaginary part, this time of the form
\[
q^2 \rightarrow q^2 + \frac{i \Delta}{\lambda c^2},
\]
where $\lambda$ is the attenuation coefficient. 

A first study shows that both effects, the pionium lifetime as well
as the photon lifetime, are roughly equally important. Whereas
these effects can result in changes of the cross sections of the order of a
few percent for higher atomic excitation energies, the target
inelastic contribution itself constitutes only a small
part of the total cross section. Therefore we conclude that the
regularisation of the cross section, even though formally needed to
get a finite result, does not lead to a change of the cross section at
the required accuracy.

%% file: juge.tex
\begin{center}
{\large\bf Hadronic Scattering in Lattice QCD}\\[0.5cm]
K.J.~Juge\\[0.3cm]
School of Mathematics, Trinity College, Dublin 2, Ireland
\end{center}

Lattice QCD calculations of the I=2 $\pi\pi$ scattering length was reviewed. Discussions of the various systematic uncertainties involved in lattice calculations (\cite{JLQCD}-\cite{Gattringer}) were given centered around the most recent $N_f=2$ simulations of the CP-PACS collaboration \cite{full}. A peculiar extrapolation to the physical pion mass limit (which has also been noted by other collaborations) were discussed using the data from the BGR collaboration who use a chirally symmetric lattice action \cite{lat03}. The extrapolations are shown in the figures below along with the result from chiral perturbation theory \cite{Colangelo}. 

\begin{minipage}[t]{8.7cm}
\hspace*{-10mm}\epsfysize=85mm\epsfbox{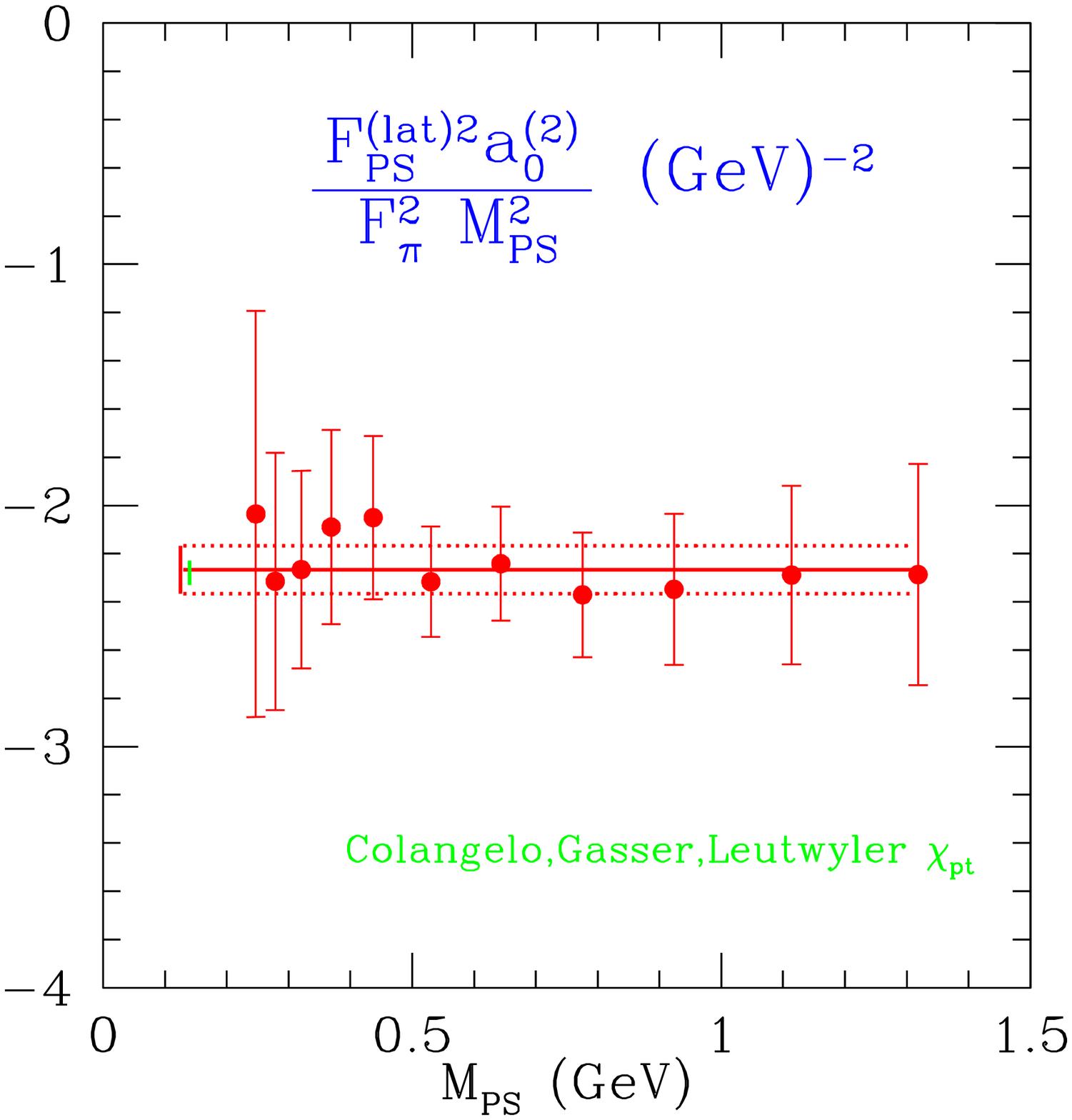}
\label{fig:mystery}
\end{minipage}
\begin{minipage}[t]{8.7cm}
\hspace*{-5mm}\epsfysize=85mm\epsfbox{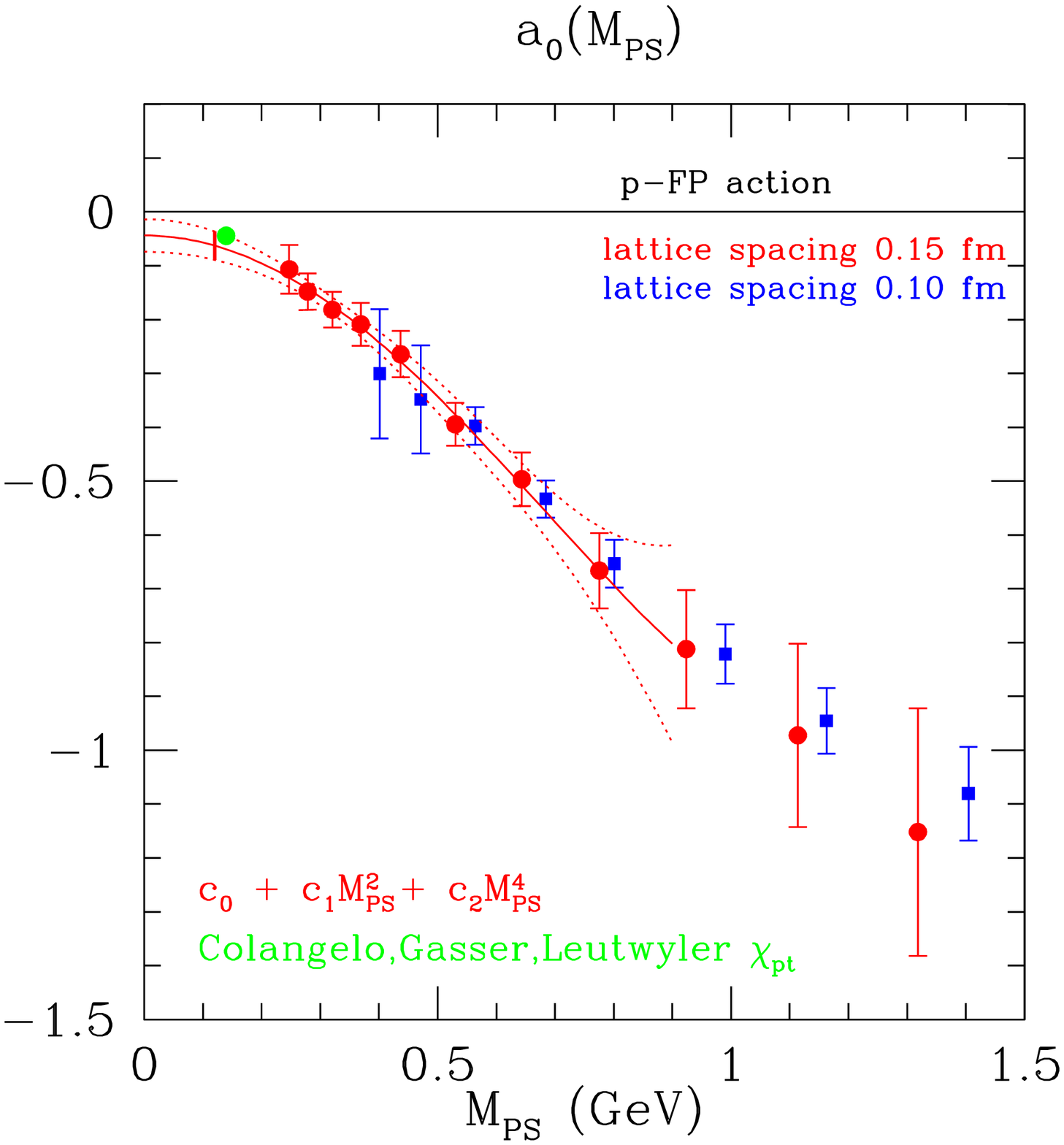}
\label{fig:extrapolation}
\end{minipage}

%% file: gotta.tex
\begin{center}
{\large\bf The pionic hydrogen experiment at PSI}\\[0.5cm]
D. Gotta, for the PIONIC HYDROGEN collaboration\\[0.3cm]
Institut f\"ur Kernphysik, Forschungszentrum J\"ulich, D-52425 J\"ulich, Germany\\[0.1cm]
\end{center}

Experiment R--98.01\,\cite{R98.01} performed at PSI aims at a high
precision determination of the strong--interaction shift
$\epsilon_{1s}$ and width $\Gamma_{1s}$ by means of X--ray
spectroscopy of ground--state transitions in pionic hydrogen. From
these quantities the $\pi$N scattering lengths and the $\pi$N coupling
constant are extracted\,\citefour{Lyu00}{Gas03}{Zemp04}{Eri02}.

Varying the target density, the observed non--effect on the
ground--state transition energy excludes any influence of cascade
effects within the experiment's accuracy yielding $\epsilon_{1s}
=7.120\pm 0.013$\,eV\,\citethree{Hen03}{Hadatom03}{meson04}.  In order to
improve significantly on the accuracy achieved for $\Gamma_{1s}$ in
previous measurements\,\cite{Sch01} three different X--ray
transitions, $\pi H(2p-1s)$, $\pi H(3p-1s)$ and $\pi H(4p-1s)$, have
been measured. The degree of Doppler broadening originating from
Coulomb de--excitation depends more strongly on the transition than
expected. In addition, the $\mu H(3p-1s)$ transition was measured,
where strong--interaction broadening is absent.

The extraction of $\Gamma_{1s}$ requires a sufficient quantitative
understanding of Coulomb de--excitation and an accurate knowledge of
the resolution of the crystal spectrometer. The determination of the
spectrometer response by using X--rays emitted from an
electron--cyclotron resonance ion trap\,\cite{ECRIT} has been
addressed by L.~Simons\,\cite{Hadatom03}. Cascade processes are
discussed by T.~Jensen\,\citetwo{Hadatom03}{Jensen05}.

From a first analysis of the Doppler broadening in pionic and muonic
hydrogen a consistent picture is obtained. In particular, the values
extracted for $\Gamma_{1s}$ for the three pionic transitions agree
within the errors.  Averaging yields a (preliminary) value of
$\Gamma_{1s} =785\pm 27$\,meV.

%% file: rusetsky.tex
\begin{center}
{\large\bf The $\pi N$ scattering lengths 
from pionic deuterium}\\[0.5cm]
Ulf-G. Mei{\ss}ner$^{1,2}$, Udit Raha$^1$ and {\bf Akaki Rusetsky}$^{1,3}$\\[0.3cm]
$^1$Universit\"{a}t Bonn, Helmholtz-Institut f\"{u}r
Strahlen- und Kernphysik (Theorie),\\
Nu\ss allee 14-16, D-53115 Bonn, Germany
\\[0.3cm]
$^2$Forschungszentrum J\"{u}lich, Institut f\" {u}r Kenphysik (Theorie),\\
D-52425 J\"{u}lich, Germany
\\[0.3cm]
$^3$On leave of absence from: High Energy Physics Institute,\\
Tbilisi State University, University St.~9, 380086
Tbilisi, Georgia
\\[0.3cm]
\end{center}

We use the framework of effective field theories  to discuss the determination
of the $S$-wave $\pi N$ scattering lengths $a_+$ and $a_-$ from the recent 
high-precision measurements of pionic deuterium observables~\cite{PSI}.
In particular, we investigate the accuracy limits on the calculations of
the pion-deuteron scattering length $a_{\pi d}$ in the multiple-scattering
series, which are derived in effective field theories. 
It has been shown that the most stringent limit is set by
the presence of the 3-particle low-energy constant (LEC) which encodes
the information about the underlying QCD dynamics at a scale 
$\simeq 1~{\rm GeV}$. The ways to estimate this constant from fitting to the
experimental data, as well as using resonance saturation, have been
discussed. We further study in detail the differences between various
field-theoretical approaches to the pion-deuteron 
problem~\citethree{Meissner}{Borasoy}{Beane}. After clarifying 
physical reasons for these (seeming) differences, they have been finally
removed from the pion-deuteron data analysis.

The results contained in this talk have been recently published in 
Ref.~\cite{deuteron}.

%% file: sainio.tex
\begin{center}
{\large\bf Progress report in $\pi$N analysis}\\[0.5cm]
M.E. Sainio\\[0.3cm]
Helsinki Institute of Physics, P.O. Box 64, 00014 University of Helsinki, Finland
\end{center}

There are still quite a few publications appearing in the literature reporting
results of pion-nucleon scattering experiments eventhough most of the meson 
factory experiments have finished their data taking. 
An exception is the pionic hydrogen experiment which is expected to improve
the level shift and width measurements. In 2004 two data analyses have been
published, the GWU partial wave analysis \cite{Arndt} and the Bugg analysis
\cite{Bugg}. They agree on the value of the pion-nucleon coupling constant
producing a number near $f^2 = 0.076$ with an error bar of about 1 \%.
A slightly higher value, $0.078$, is obtained by Ericson et al. \cite{Ericson}
in an analysis of the pionic hydrogen data.

The standard for pion-nucleon analysis was set by the work of the Karslruhe
group in the 70's and 80's. In an attempt to update that analysis, we have
in Helsinki been working on the expansion method for implementing the fixed-$t$
dispersion relations. 
As an example for the $C^+$ amplitude ($C=A+\nu/(1-t/4m^2)B$) we have
\begin{eqnarray}
C^+(\nu,t) = C^+_N(\nu,t) + H(Z,t)\sum_{n=0}^N c^+_n Z^n,\nonumber
\end{eqnarray}
where $H$ is adjusted to the asymptotic behaviour of
the amplitude and
\begin{eqnarray}
Z(\nu^2,t) = \frac{\alpha - \sqrt{\nu_{th}^2-\nu^2}}
                   {\alpha + \sqrt{\nu_{th}^2-\nu^2}}.
\nonumber
\end{eqnarray}
Currently we are still working on data generated
with an existing phase shift solution (KA84), but, gradually, the 
updated data base will be used. In the forward direction the expansion method
is simple to use. For the $C^{\pm}$ amplitudes input for the imaginary
parts (total cross sections) and real parts (3 data points for Re$D^+(\omega,t=0)$)
yield an adequate reproduction of the Karlsruhe forward amplitudes, see
\cite{MES}.

%% file: zemp.tex
\begin{center}
{\large\bf Decay width of the ground state of pionic hydrogen \\in QCD + QED}\\[0.3cm]
P. Zemp\\[0.2cm]
ITP, University of Bern, Sidlerstrasse 5, 3012 Bern, Switzerland\\[0.1cm]
\end{center}
The width and the shift of the ground state of pionic hydrogen have been
calculated in the framework of potential models in ref.~\cite{potential}.
Later the energy shift has been evaluated in a systematic low-energy expansion
in~\cite{Lyubovitskij} whereas the width was considered in the same framework
in~\cite{Zemp}. Counting $\alpha \sim 1/137$ and $m_d-m_u$ as small quantities
of order~$\delta$, the result reads 
\begin{gather}
  \label{eq:general}
  \Gamma_{1{\rm s}} = 8\,\alpha^3 M_r^2 \; p^\star_1 \; \left ( 1+\frac{1}{P}
  \right )\;[a_{0+}^-(1+\delta_\Gamma)]^2+ \mathcal{O}(\delta^{5})\;,
\end{gather}
with the reduced mass $M_r = M_{\pi^{\scriptscriptstyle +}} m_p/
(M_{\pi^{\scriptscriptstyle +}} + m_p)$,
the Panofsky ratio $P$ and the isospin odd scattering length $a_{0+}^-$. The
momentum $p^\star_1$ is the center-of-mass momentum of the process
$\pi^-p\to\pi^0n$. The isospin correction term $\delta_\Gamma$ is 
\begin{gather}
  \delta_\Gamma = \frac{\epsilon}{a_{0+}^-} + K + \delta_{\epsilon}^{\rm
    vac}\;,\\
  K = 4M_r\,\alpha\,(1-\ln{\alpha})(a_{0+}^+ + a_{0+}^-) +2M_r(M_\Sigma -\bar
  M_\Sigma)(a_{0+}^+)^2\; + o\,(\delta)\;.
\end{gather}
Vaccum polarization effects \cite{Eiras} are contained in
$\delta_\epsilon^{\rm vac}$. The general expression~(\ref{eq:general}) agrees
in its form with the one derived by Sigg {\it et al.} \cite{potential} up to a
slight difference in the definition of the kinematic prefactor $p_1^\star$.
The quantity $\epsilon$ can be calculated in CHPT. At tree level
calculation~\cite{Zemp} yields
\begin{gather}
  \delta_\Gamma =(0.6\pm 0.2)\times 10^{-2}\,\,\,[\mbox{CHPT, leading
    order}]\;.
\end{gather}
This result can be compared with $\delta_\Gamma=-(1.3\pm 0.5)\times 10^{-2}$
found by Sigg {\it et al.} \cite{potential}. A one-loop evaluation of
$\epsilon$ is underway \cite{Buettiker}. For further investigations, see
ref.~\cite{Ericson}.

%% file: ericson.tex
\begin{center}
{\large\bf  The $\pi ^-p$ atomic level shift and  Isospin Violation in $\pi N$ Scattering}\\[0.5cm]
 {\bf T. E. O. Ericson}$^1$ and A. N. Ivanov$^2$\\[0.3cm]
$^1$Theory Division, Physics Department, CERN, 
CH-1211 Geneva 23, Switzerland\\[0.1cm]
$^2$ Atomic Institute of the Austrian Universities, Vienna
    University of Technology, \\  A-1040 Wien and 
    SMI of the Austrian Academy of Sciences, A-1090,
    Wien\\[0.1cm]
\end{center}

The high precision of data from pionic hydrogen and deuterium permit in principle the 
determination of the corresponding $\pi N $ scattering lengths to
 a precision of up to $0.2\%$ and the e.~m. contributions  become  essential.  It is important to have a 
clear physical picture of their origin and uncertainties. As a complement to 
the approach via effective field theory with QCD+QED \cite{JG02} we study the 
e.~m. terms which are generated by the strong interaction,
 irrespective of its origin. Our starting point is the empirical $\pi N$ low energy
  expansion  in the absence of
the external Coulomb field, which may  have intrinsic
 isospin violating terms. We assume  that the e.~m. form factors of
 the pion and nucleon are empirically known  and so are the axial form factors.
 All masses are taken to have the physical values.
In short, we have no free parameters. 
The starting point is the Coulomb problem for the extended charge distribution.
The strong interaction at short range is then a  perturbation  in the spirit 
of a pseudopotential. By minimal e.~m. coupling, or equivalently, 
 interaction energy the $\pi N$ scattering amplitude has its
 energy  shifted by the  Coulomb energy at the origin; this is shown in a model \cite{ERI04}.
 The wave function at the origin is the one of the extended 
charge distribution and not the usually used  Bohr wave function.
In addition a characteristic cusp factor appears to second order in the scattering length,
 but it is insensitive to assumptions. 

All these corrections are of the order of 1\% or less with little uncertainty and they are stable with respect to assumptions.
The more general case case of an extended strong interaction region leads  only to
 weak modifications in practice.

At threshold and in the heavy baryon limit the additional e.~m. mechanisms are greatly simplified.
Only the dispersive counterpart $\pi N\to (N,\Delta )\gamma \to \pi N$ of 
the Kroll-Ruderman radiative capture process contributes with an axial matrix element and no free parameter
 \cite{ERI05}. 
  Already the $N\gamma $ intermediate state is an  important contribution to the $\pi ^-p$ amplitude, 
which  increases by
 9.3\%  when the intermediate $\Delta \gamma $ states are included degenerate with the nucleon. The isovector amplitude 
 in $m_{\pi }\ln m_{\pi }$ agrees with  Gasser et al. \cite{JG02} using only 
$N\gamma $ intermediate states. This term is nearly  canceled by the
 $\Delta \gamma $ contribution in the limit of  the $N\Delta $ mass splitting $\omega _{\Delta } =0$. 
For $\omega _{\Delta } \neq 0$ the present mechanism contributes $6.5\%$ to the $\pi ^-p$ 1s level shift. 
 These radiative processes 
are closely governed by the the same $\Delta N$ physics  as  
 in the static limit of $\pi N$ scattering \cite{TE88}.

%% file: jensen.tex
\begin{center}
{\large\bf Kinetic energy distributions in pionic hydrogen}\\[0.5cm]
T.S. Jensen\\[0.3cm]
LKB, Ecole Normale Sup\'erieure et Universit\'e Pierre et Marie Curie,
Case 74,\\
 4 Place Jussieu,
F-75252 Paris Cedex 05,
France
\end{center}

The goal of the new pionic hydrogen experiment at the
Paul--Scherrer-Institut~\cite{psiexp} is to measure the line shapes of
the $np\to1s$ X--ray transitions in $\pi^-p$ with high precision and
to obtain the strong interaction $1s$ shift and width from the
results. A difficult problem arises in the case of the extraction of
the $1s$ width: the $\pi^-p$ atoms are in motion at the moment of the
radiative transitions so the observed line profiles are broadened due
to the Doppler effect.  Highly energetic (tens of electron--Volts)
$\pi^-p$ atoms can be produced through Coulomb
deexcitation~\cite{bracci78}
\[ (\pi^-p)_{n_1}+\mathrm{H}_2\to (\pi^-p)_{n_2}+\mathrm{H}+\mathrm{H}\; \]
or  by the break--up of excited molecular ion states~\cite{kilic04}
\[ (\pi^-pp)^+_{\nu J}\to(\pi^-p)_{n_2}+p\;.\]
Presently, the quantitative understanding of these processes do not
allow us to make {\it ab initio} cascade model
predictions~\citetwo{jensen02}{jensen03} which are sufficiently
accurate for the new experiment.  These difficulties have led to the
suggestion of an alternative approach for extracting the $1s$
width~\cite{jensen04}. The idea consists of combining reliable cascade
model input with a fitting procedure which uses parameters to describe
the poorly known kinetic energy distributions. Reliable calculations
of cross sections are possible for the highly energetic $\pi^-p$ atoms
because the molecular structure of the target hydrogen molecule is
less important than for lower energies. This means that the {\it
  relative} contributions of for example the $4\to3$ Coulomb component
at 73 eV to the broadening of the $3p\to1s$ and $2p\to1s$ X--ray lines
can be predicted.  By using these results, a more accurate value of
the strong interaction width can be obtained as compared to a fit
without constraints.  \vspace{0.2cm}

\noindent  Acknowledgment: This work was supported by the Swiss National Science Foundation.

%% file: pineda.tex
\begin{center}
{\large\bf The chiral structure of the (muonic) hydrogen}\\[0.5cm]
A. Pineda$^1$\\[0.3cm]
$^1$Dept. d'Estructura i Constituents de la Mat\`eria, U. Barcelona,\\
Diagonal 647, E-08028 Barcelona, Catalonia,  Spain
\end{center}

High precision measurements in atomic physics provide with a unique
place to fix some hadronic parameters related with the proton
elastic and inelastic electromagnetic form factors, like the proton
radius, magnetic moment, polarization effects, etc.... One
complication in this program comes from the fact that widely separated
scales are involved in these physical processes. Therefore, it becomes
important to relate the physics at these disparate scales in a model
independent way. Effective field theories (EFT's) are a natural
approach to this problem.  In particular, we need an EFT 
at atomic scales, a natural candidate for which is potential NRQED \cite{pNRQED}.
By relating this theory with HBET \cite{HBET}, it is possible to obtain the chiral 
(and model independent) structure of the level splittings of 
(muonic) hydrogen. Over the last years this idea has been applied 
in \cite{Pineda:2002as}. The matching between HBET 
coupled to photons and leptons and the relevant
EFT at atomic scales have been performed. This matching can be
organized in a perturbative expansion in $\alpha$, $1/m_p$ and the chiral
counting. We then computed the $O(m_{l}^3\alpha^5/m_p^2\times (\ln m_{\pi}))$
contribution to the Hyperfine splitting (as well as other large logs: 
$\ln \Delta$, ...) and compared with experiment. This contribution 
can explain about 2/3 of the difference between experiment and the 
pure QED prediction when setting the renormalization scale at the 
$\rho$ mass. The difference can be used to give an estimate of 
the matching coefficient of the 
spin-dependent proton-lepton operator in HBET.

The definition of the electromagnetic proton radius has also been studied. 
It happens to be ambiguous once 
electromagnetic corrections are considered. This is relevant in view of the possibility
to measure the muonic hydrogen Lamb shift at the 30 ppm level at the PSI 
\cite{Taqqu} and,
consequently, to obtain the proton radius with an improved accuracy of, at
least, one order of magnitude better. We have argued that a natural 
definition can be given within an EFT framework 
in terms of a matching coefficient. The definition of the neutron radius 
was also discussed. The hadronic corrections to the lamb shift 
(for the polarizability effects only with logarithmic accuracy) were computed 
within HBET up to 
$O(m_{l}^3\alpha^5/m_p^2\times F(m_l/m_{\pi}))$. 
It was found that they diverge in the inverse 
of the pion mass in the chiral limit. Moreover, at this order, 
they are a prediction of HBET 
since unknown counterterms are subleading.

%% file: marton.tex
\begin{center}
{\large\bf Precision Measurements with Kaonic Atoms}\\[0.5cm]
{\bf J. Marton} for the DEAR and SIDDHARTA Collaborations\\[0.3cm]
Stefan Meyer Institut, Austrian Academy of Sciences,\\[0.1cm]
Boltzmanngasse 3, 1090 Wien, Austria
\end{center}

The spectroscopy of X-rays from exotic atoms like pionic and
kaonic hydrogen atoms provides precision studies of strong
interaction. Kaonic atoms are produced at the DA$\Phi$NE
electron-positron collider in Frascati. The X-ray transitions in
these exotic atoms are studied using an array of CCD X-ray
detectors. We obtained the following results:
\\[0.1cm]
1) \textit{Kaonic nitrogen:} We measured the X-ray lines and
yields of the 7-6, 6-5 and 5-4 transitions for the first time
\cite{Ishiwatari04}. Kaonic nitrogen is found to be nearly
completely stripped of electrons, thus allowing a new kaon mass
measurement with a high resolution X-ray detection. \\[0.1cm]
2) \textit{Kaonic hydrogen:} The X-ray spectrum of the kaonic
hydrogen K transitions was extracted. The values for the strong
interaction shift $\epsilon_{1s} = -193 \pm 37 (stat.) \pm 6
(syst.) eV$ and width $\Gamma_{1s} = 249 \pm 111(stat.) \pm 30
(syst.) eV$ were extracted \cite{pubH}. \\
The repulsive kaon-proton interaction shown in \cite{iwa} is
verified, but we find smaller values for the shift and width with
significantly smaller errors.

\begin{figure}[h]
\begin{center}
\includegraphics[width=6.cm]{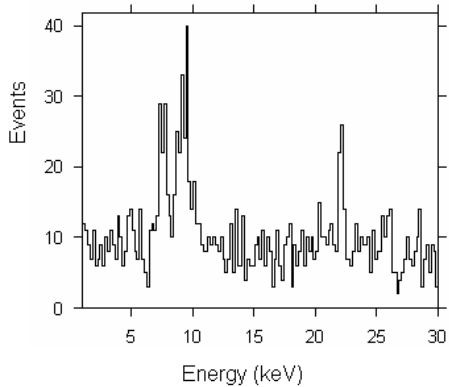}
\end{center}
\caption{Monte Carlo simulated X-ray spectrum of kaonic deuterium
using SDDs.} \label{fig1}
\end{figure}

Our experimental program is proceeding toward high precision
measurements with new large area X-ray detectors providing timing
capability (SDDs) presently in development \footnote{EU 6$^{th}$
Framework Program, I3-HP Contract No. RII3-CT-2004-506078}.

%% file: raha.tex
\begin{center}{\Large{\bf Spectrum and decays of the Kaonic Hydrogen
}}\\
\bigskip
{U.-G. Mei\ss ner, {\bf U. Raha}, and A. Rusetsky}\\
\bigskip
{\em Helmholtz-Institut f\"ur Strahlen- und Kernphysik, Universit\"at Bonn}\\
{\em Nu\ss allee 14-16, 53115 Bonn, Germany}
\end{center}

Recent accurate measurements \citetwo{Baldini}{Cargnelli} of the strong
energy shift and the lifetime of the ground state of kaonic hydrogen
by DEAR collaboration at LNF-INFN allow one to extract the precise
values of the $KN$ scattering lengths from the data. To this end, one
needs to relate the latter quantities to the observables of the kaonic
hydrogen at the accuracy that matches the experimental precision.  In
our recent investigations, the problem is considered within the
non-relativistic effective Lagrangian approach, which has been
previously used to describe the bound $\pi^+\pi^-$ , $\pi^-p$ and $\pi
K$ systems (see, e.g.  \citefour{Gall}{Lyubovitskij}{Zemp}{Schweizer}). We
obtain \cite{Meissner} a general expression of the strong shift of the
level energy and the decay width in terms of the $KN$ scattering
lengths, at $\mathcal{O}(\alpha,m_d-m_u)$ as compared to the
leading-order result. It is shown that, due to the presence of the
unitarity cusp in the $K^-p$ elastic scattering amplitude above
threshold, the isospin-breaking corrections turn out to be very large.
This, however, does not affect the accuracy of the extraction of the
scattering lengths from the experiment.

%% file: ivanov.tex
\begin{center} 
{\large\bf On the energy level displacement of the
    ground state of kaonic deuterium}\\[0.5cm] 
{\bf A. N. Ivanov}, M.  Cargnelli, M. Faber, H. Fuhrmann, V. A.
  Ivanova,\\ J. Marton, N. I.  Troitskaya, and J. Zmeskal\\[0.3cm]
  Atomic Institute, Vienna University of Technology, Stefan Meyer
  Institute, Austrian Academy of Sciences, Vienna, Austria and State
  Polytechnic University of St. Petersburg,
  St. Petersburg, Russia\\[0.3cm]
\end{center}

The energy level displacement of the ground state of kaonic deuterium
is defined by the DGBTT formula\,\footnote{This is the abbreviation of
  the Deser--Goldberger--Baumann--Thirring--Trueman formula.}
\begin{eqnarray}\label{label1}
- \epsilon_{1s} + i\,\frac{\Gamma_{1s}}{2} = 2\alpha^3\mu^2\,
f^{K^-d}_0(0) = 602\,f^{K^-d}_0(0),
\end{eqnarray}
where $\alpha = 1/137.036$ is the fine--structure constant, $\mu =
m_Km_d/(m_k + m_d) = 391\,{\rm MeV}$ is the reduced mass of the $K^-d$
pair for $m_K = 494\,{\rm MeV}$ and $m_d = 1876\,{\rm MeV}$ and
$f^{K^-d}_0(0)$ is the S--wave amplitude of $K^-d$ scattering at
threshold. We show Ref.\cite{IV4} that the S--wave scattering length
${\cal R}e\,f^{K^-d}_0(0) = a^{K^-d}_0$ is described well by the
Ericson--Weise--like scattering length of $K^-d$ scattering
Ref.\cite{TE88}. The imaginary part ${\cal I}m\,f^{K^-d}_0(0)$ we
calculate in terms of the contributions of the two--body reactions
$K^-d \to NY$, where $NY = n\Lambda^0$, $n\Sigma^0$ and $p\Sigma^-$,
and the experimental data on the two--body reaction rates (see
Ref.[33] of Ref. \cite{IV4}).  Our theoretical predictions for the
two--body reaction rates agree well with the experimental data.  The
S-wave amplitude of $K^-d$ scattering at threshold is equal to
$f^{K^-d}_0(0) = (-\,0.540 \pm 0.095) + i\,(0.521 \pm 0.075)\, {\rm
  fm}$ Ref.\cite{IV4}.  Our prediction for the energy level
displacement of the ground state of kaonic deuterium Ref.\cite{IV4}
\begin{eqnarray}\label{label3}
-\,\epsilon_{1s} + i\,\frac{\Gamma_{1s}}{2} = 602\,f^{K^-d}_0(0) = 
 (-\,325 \pm 60) + i\,(315 \pm 50)\;{\rm eV}
\end{eqnarray}
can be used for the planning experiments by the DEAR/SIDDHARTA
Collaborations at Frascati (C.  Guaraldo {\it et al.}, {\it
  Measurement of kaonic hydrogen with DEAR at DAFNE and future
  perspectives}, Naples, Italy, November 3--7, 2004 and the report by
J.  Marton ``{\it Precision measurements with kaonic atoms}'' at this
Workshop {\it HadAtom05}).

%% file: lyubovitskij.tex
\begin{center}
{\large\bf Nucleon stucture in Lorentz covariant chiral quark model}\\[0.5cm]
Amand\ Faessler, Th.\ Gutsche, V.\ E.\ Lyubovitskij, K. \ Pumsa-ard\\[0.3cm]
Institut f\"ur Theoretische Physik, Universit\"at T\"ubingen,\\
Auf der Morgenstelle 14, D-72076 T\"ubingen, Germany 
\end{center}

We developed a manifestly Lorentz covariant chiral quark model for the 
study of baryons as bound states of constituent quarks. The approach is 
based on a non-linear chirally symmetric 
Lagrangian~\cite{Gasser:1987rb}-\cite{Kubis:2000zd}, which involves 
effective degrees of freedom - constituent quarks and the chiral (meson) 
fields. In a first step, this Lagrangian can be used to perform a dressing 
of the constituent quarks by a cloud of light pseudoscalar mesons and other 
heavy states using the infrared dimensional regularization of loop 
diagrams~\cite{Becher:1999he}.      
We calculate the dressed transition operators with a proper 
chiral expansion which are relevant for the interaction of quarks with 
external fields in the presence of a virtual meson cloud. 
These operators are used in the calculation of baryon matrix 
elements~\cite{PCQM}.  

Our main result is: we perform a model-independent factorization of the 
effects of hadronization and confinement contained in the matrix elements 
of the bare quark operators and the effects dictated by chiral symmetry. 
The calculation of these effects can be done 
independently. In particular, the computing of matrix elements of bare 
quark operators can then be relegated to quark models based on specific 
assumptions about hadronization and confinement. 
The nucleon form factors have a correct infrared-singular 
structure dictated by chiral symmetry~\citetwo{Kubis:2000zd}{Beg:1973sc}. 
We reproduce the leading nonanalytic (LNA) contributions to the magnetic 
moments and the charge and magnetic radii of the nucleons. 
The LNA contribution to the magnetic 
moments is proportional to the pseudoscalar meson mass $M_P$. 
The nucleon radii are divergent in the chiral limit.   
The LNA contribution to the charge radii is proportional to the 
chiral logarithm ${\rm ln}(m_N^2/M_P^2)$. The LNA contributions 
to the magnetic radii are represented by the same logarithm as in 
the case of the charge radii and by the singular term proportional to 
$1/M_P$.  

This work was supported by the DFG under contracts FA67/25-3 and GRK683. 
This research is also part of the EU Integrated Infrastructure Initiative 
Hadronphysics project under contract number RII3-CT-2004-506078.

%% file: mannelli.tex
\begin{center}
{\large\bf 
Study of pion-pion interaction in $K \to 3 \pi$ decays} \\[0.5cm]
Italo MANNELLI\\[0.3cm]
Scuola Normale Superiore
PISA - Italy
\end{center}

The NA48/2 experiment at the CERN SPS is searching for direct CP 
cviolation in K+- decays to three pions. The experiment uses simultaneous 
K+ and K- beams of $60\pm3$ GeV/c momentum which overlap in the fiducial 
region. During 2003 and 2004 about 4 billions fully reconstructed 
$K^\pm \to \pi^\pm \pi^+ \pi^-$ decays and more than 100 millions $K^\pm
\to \pi^\pm \pi^0 \pi^0$ have been collected.
The results from a study of the $\pi^0 \pi^0$ invariant mass distribution
$M_{00}$  in a partial sample of 30 millions decays $K^\pm \to \pi^\pm
\pi^0 \pi^0$ are reported, showing a cusp-like structure centerd at $M_{00} =
2 M_{\pi^+}$. The experimental apparatus consists mainly of a magnetic
spectrometer, with a momentum resolution $\sigma(p)/p = 1.02\% (+) 0.044\%$
and a liquid Krypton calorimeter segmented transversally into 13248
2cm$\times$2cm cells. 
The calorimeter is 27 $X_0$ thick and has and energy resolution for photons
$\sigma(E)/E = 0.032/\sqrt{E} (+) 0.09/E (+) 0.0042$ ($E$ in GeV) and space
resolution $\sigma(x)=\sigma(y) = 0.42/\sqrt{E} (+) 0.06$ cm thus allowing
the accurate detection of multi-photons events. 
By imposing the $\pi^0$ invariant mass to pairs of photons and requiring
the compatibility of a common origin for four-photon events grouped in
two $\pi^0$ pairs and at the same time the Kaon invariant mass including
the additional charged pion detected in the magnetic spectrometer, decays
of $K^\pm \to \pi^\pm \pi^0 \pi^0$ can be fully reconstructed with negligible 
background.
The resolution in the $\pi^0 \pi^0$ invariant mass near $M_{00} = 279$ MeV is
typically $0.5$ MeV.  
The distribution of the events as a function of $M_{00}^2$ shows a cusp like
structure around $M_{00}^2 = 4 M^2_{\pi^+}$ which is definitely not due to an 
instrumental effect, given the intrinsic smoothness of the acceptance.
It is impossible to fit the distribution with the usual parametrization
employed by the PDG whenever the cusp region is included in the range
for the fit.
A good fit, with $\chi^2 = 149/147$ d.o.f., is obtained by using the 
expression of the matrix element as specified in \cite{CI}
which takes into account one and two-loops pion-pion 
scattering, in particular $\pi^+ \pi^- \to \pi^0 \pi^0$ charge exchange, and 
allowing for the contribution of the formation of pionium atoms, with 
their subsequent $\pi^0\pi^0$ decay, with a branching ratio as calculated by 
Silagadze \cite{Silagadze}.
The main ingredient of the fit is the difference  $a^0_0-a^2_0$  between
the I=0 and I=2 S-wave pion-pion scattering lenghts. With the present
sample the statistical error in the determination of $a^0_0-a^2_0$ is about
4\% while the systematic uncertainty is still under evaluation.

In conclusion the study of the $M_{00}^2$ distribution in $K^\pm \to
\pi^\pm \pi^0 \pi^0$ decays by the NA48/2 experiment allows for a novel and
potentially more precise determination of $a^0_0-a^2_0$ which is made
possible in particular by the quality of its multi-photon detection
capability.

%% file: tarasov.tex
\begin{center}
{\large\bf The unitary correction to the Moli\`ere--Fano
  multiple scattering theory}\\[0.5cm]
S. Bakmaev, {\bf A. Tarasov}, and O. Voskresenskaya\\[0.3cm]
Joint Institute for Nuclear Research, 141980 Dubna,Moscow Region, Russia\\[0.1cm]
\end{center}

The Moli\`ere--Fano multiple scattering theory
\citetwo{Moliere}{Fano} is the most used tool for the accounting of
the charged particles' multiple scattering effects at experimental
data processing. Estimation of the theory accuracy is of especial
importance in the case of DIRAC experiment for it's high angular
resolution. One possible source of the inaccuracy of M--F theory is
the usage in \cite{Moliere} of the approximate expression for the
amplitude of elastic scattering of charged particles by atoms, which
violates the optical theorem (unitarity condition). We have estimated
the relative corrections the the parameters of M--F theory, resulting
from restoring of unitarity in the particle--atom scattering theory
and found that they are of order $Z\alpha^2$, where $Z$ is the atomic
number of target atoms.

%% file: voskr.tex
\begin{center}
{\large\bf Production of pioniums in the coherent nuclear-nuclear collisions}\\[0.5cm]
S. Bakmaev  and O. Voskresenskaya\\[0.3cm]
Joint Institute for Nuclear Research, 141980 Dubna,Moscow Region, Russia\\[0.1cm]
\end{center}

We conside the production of $nP$-states of pionium atoms in the
coherent collisions of a projectile $A_p$ and target $A_n$ nuclei 
\begin{equation}
  \label{vs:eq:1}
  A_p + A_t \to A_p + A_t + A_{2\pi}(nP)
\end{equation}
as the possible source of $A_{2\pi}(nP)$ beam for the pionium
Lamb-shift measurements \cite{Nem:HA03}.

Since the quantum numbers of pionium in $nP$-states are the same as
photon's ones the coherent photoproduction process 
\begin{equation}
  \label{vs:eq:2}
  \gamma + A_t \to A_{2\pi}(nP) + A_t 
\end{equation}
is quite intence in GeV-region. Another advantage of the coherent
reaction (\ref{vs:eq:2}) is the sharp angular distribution of produced
pioniums.

The source of ``effective'' photons could be provided
by projectile nuclear $A_p$. For the crude estimation of
$A_{2\pi}(nP)$  yeilds the following formula can be used.
\begin{equation}
  \label{vs:eq:3}
  \omega \frac{d\sigma}{d\omega} = n_{eff}(Z_p,\omega) \alpha^6
  (\sigma^{tot}_{\pi A_t} m_{\pi})^2 \left(\frac{1}{n^3}- \frac{1}{n^5}\right) 
\end{equation}
Here $n_{eff}$ is the number of ``effective'' photons with the energy
$\omega=E(A_{2\pi})$ produced by projectile nuclear with charge $Z_p$
\cite{vosk:bert} and $n$ is the principal quantum number of pionium in
the $nP$-state.

%% file: friedman.tex
\begin{center}
{\large\bf Pion-nucleus interaction at low energy}\\[0.5cm]
{\bf E. Friedman}
\footnote{Supported by the
Israel Science Foundation grant No. 131/01.}
\\[0.3cm]
Racah Institute of Physics, The Hebrew University, Jerusalem,
Israel\\[0.1cm]
\end{center}

An over-view of the pion-nucleus optical potential, as obtained from
global fits to large data sets encompassing the whole of the periodic table
is presented. Data bases of well over 100 points have existed since the
1990's, and some 10 points due to `deeply bound' pionic atom states have
been added in recent years.
Interest  in the pion-nucleus interaction
at low energies
has been focused  on the $s$-wave part of the pion-nucleus
optical potential, where the so-called `anomalous' repulsion of the 
$s$-wave pionic atom potential is the empirical finding
that the strength of the repulsive $s$-wave
potential inside nuclei is nearly double the values expected on
the basis of the free $\pi N$ interaction.
Large scale fits to pionic atom data showed \cite{Fri02} 
that the model by  Weise \cite{Wei01}
where the {\it isovector} scattering amplitude becomes density dependent,
due to a chiral-motivated density dependence of the pion decay 
constant, is capable 
of removing the anomaly. 
In an alternative approach the empirical energy dependence of 
the {\it isoscalar} amplitude, within 
the minimal substitution requirement \cite{ETa82} of
$E \to E - V_{c}$, where $V_{c}$ is the Coulomb potential, was also found
\cite{FGa04} to remove the anomaly. 
Both models yield equally good fits to the data.

The study of the pion-nucleus interaction was extended above threshold
in  recent precision measurements of the elastic scattering of 21 MeV
$\pi ^\pm$ by several nuclei \cite{FBB04}.
The purpose was to see
if the  anomaly is observed also above threshold.
In the scattering scenario, unlike in the atomic
case, one can study both charge states of the pion, thus increasing
sensitivities to isovector effects and to the energy dependence 
of the isoscalar amplitude due to the Coulomb interaction.
Elastic scattering experiments were performed on targets of Si, Ca, Ni
and Zr and reasonably good fits to the data were obtained with the 
conventional pion-nucleus potential, which again displayed the `anomalous'
repulsion. Greatly improved fits were obtained when the energy dependence
was included, but some of the anomaly still remained. With the density
dependence mentioned above the anomaly disappears, and including also the
energy dependence the fits are very good and the anomaly is absent.
Thus from the quality of fits the energy dependence is required by the
data, and the anomaly is removed if the chiral-motivated density dependence
is included in the model.

%% file: gasser.tex
\begin{center}
{\large\bf Concluding remarks}\\[0.5cm]
J. Gasser\\[0.3cm]
ITP, University of Bern, Sidlerstrasse 5, 3012 Bern, Switzerland
\end{center}
I discussed the present status of theory and  experiment in Goldstone 
boson scattering, and relied for this mainly on the very interesting 
talks presented at the workshop.
My personal judgment of the situation is summarized in the
table. 
\newcommand{\okt}{\underline{\underline{$\surd$}}}
\newcommand{\oko}{\underline{$\surd$}}
\newcommand{\okz}{$\surd$}
\newcommand{\bc}{\begin{center}}
\newcommand{\ec}{\end{center}}
\bc
\begin{tabular}{cc|c|ccc}
Theory&&$\hspace{1cm}\Longleftrightarrow\hspace{1cm}$&Experiment&\\ 
      &status\hspace{1cm}      &status&&status\\ \hline
\hline\\[1mm] 
\multicolumn{6}{c}{{\hspace{1.7cm}\framebox{$\pi\pi\rightarrow\pi\pi$}}}\\&&\\
scattering lengths&\okt&\oko&$K_{e4}$&\okt&\\
&&\okt&$A_{\pi\pi}$&p&\\
&&\okz&$K_{3\pi}$&p&\\[1mm]
\multicolumn{6}{c}{\hspace{1.7cm}
\framebox{$\pi N\rightarrow \pi N$}}\\&&\\
scattering lengths&\oko&\okt&$\pi N\rightarrow \pi N$ &\oko&\\
&&\oko&$A_{\pi p}$&\oko&\\
&&\okz&$A_{\pi d}$&p&\\ [2mm]
\multicolumn{6}{c}{\hspace{1.7cm}
\framebox{$\pi K\rightarrow\pi K$}}\\&&\\
scattering lengths&\oko&\okt&$A_{\pi K}$&p&\\[2mm]
\multicolumn{6}{c}{\hspace{1.7cm}
\framebox{$K p\rightarrow K p$}}\\&&\\
scattering lengths&p&\oko&$A_{K p}$&\oko &\\
&&p&$A_{K d}$&p&\\
\end{tabular}
\ec
\vskip3mm
To illustrate the table, the theoretical 
status of $\pi\pi$ scattering lengths is very good (\okt), 
the (theoretical)  relation of the scattering lengths to the 
cusp in $K_{3\pi}$ decays ($\Longleftrightarrow$) needs improvement (\okz),
 and the experimental 
status of determining scattering lengths 
from $K_{3\pi}$ decays is qualified as ``in progress'' (p).

Lattice calculations have
the potential to provide in the near future reliable information 
on the various scattering lengths from first principles.

%% file: hadatom05_proc.bbl
\bigskip\bigskip




\begin{thebibliography}{12}

\bibitem{98}
Proceedings of the International Workshop ``Hadronic Atoms and 
Positronium in the Standard Model'', Dubna, 26-31 May 1998, 
Ed. M.A. Ivanov, A.B. Arbuzov, E. A. Kuraev, V.E. Lyubovitskij,  A.G. Rusetsky.

\bibitem{99}
J.~Gasser, A.~Rusetsky and J.~Schacher,
``HadAtom99'', arXiv:hep-ph/9911339.

\bibitem{01}
J.~Gasser, A.~Rusetsky and J.~Schacher,
``HadAtom01'', arXiv:hep-ph/0112293.

\bibitem{02}
L.~Afanasyev, A.~Lanaro and J.~Schacher,
``HadAtom02'', arXiv:hep-ph/0301266.

\bibitem{03}
J.~Gasser, A.~Rusetsky and J.~Schacher,
``HadAtom03'', arXiv:hep-ph/0401204.

\end{thebibliography}

\begin{thebibliography}{99}

\bibitem{DIRAC_proposal:1994}
B.~Adeva et. al., DIRAC proposal, 
CERN-SPSLC-95-1, SPSLC-P-284 (1995).\\[-5mm]

\bibitem{ChPT_Colangelo:2001}
G.~Colangelo, J.~Gasser and H.~Leutwyler, 
Nucl. Phys. B603 (2001) 125.\\[-5mm] 

\bibitem{Nemenov:1985}
L.~Nemenov, Yad. Fiz. 41 (1985) 980; 
(Sov. J. Nucl. Phys. 41 (1985) 629).\\[-5mm]

\bibitem{atoms_interact}
L.G.~Afanasyev and A.V.~Tarasov, Phys. At. Nucl. 59 (1996) 2130;
T.A.~Heim et al., J. Phys. B33 (2000) 3583; 
see also references there.\\[-5mm]

\bibitem{DIRAC_observation:2004}
B.~Adeva et al., J. Phys. G30 (2004) 1929.\\[-10mm]


\end{thebibliography}

\vspace{-2mm}

\begin{thebibliography}{99}
  
\bibitem{dirac} B. Adeva et al., {\it Lifetime measurement of
    $\pi^+\pi^-$ atoms to test low energy QCD predictions}, Proposal
  to the SPSLC, CERN/SPSLC 95--1, SPSLC/P 284, Geneva, 1995.

\bibitem{nem85} L.L.~Nemenov, Yad.Fiz. {\bf 41} (1985) 980.
\bibitem{nem01} L.L.~Nemenov, V.D.~Ovsiannikov, Phys.Lett. {\bf B 514} (2001) 247.
\bibitem{nem02} L.L.~Nemenov, V.D.~Ovsiannikov and E.V.~Tchaplyguine,
  Nucl.Phys. {\bf A 710} (2002) 303.

\end{thebibliography}

\begin{thebibliography}{99}
\bibitem{Adeva:2004bd}
  B.~Adeva {\it et al.}  [DIRAC Collaboration],
  J.\ Phys.\ G {\bf 30} (2004) 1929
  [arXiv:hep-ex/0409053].
\bibitem{Halabuka:1999bm}
  Z.~Halabuka, T.~A.~Heim, K.~Hencken, D.~Trautmann and R.~D.~Viollier,
  Nucl.\ Phys.\ B {\bf 554} (1999) 86.
\bibitem{Heim00} T. Heim {\it et al.}, J. Phys. B {\bf 33} (2000) 3583
\bibitem{Heim01} T. Heim {\it et al.}, J. Phys. B {\bf 34} (2001) 2763
\bibitem{Schumann02} M. Schumann {\it et al.}, J. Phys. B {\bf 35} (2002)
  2683
\bibitem{Schumann03} M. Schumann, Ph. D. thesis, U Basel (2003) unpublished
\bibitem{Fano63} U. Fano, Annu. Rev. Nucl. Sci. 13 (1963) 1
\bibitem{Ginzburg95} I. F. Ginzburg, hep-ph/9509314
\bibitem{Melnikov96} K. Melnikov and V. G. Serbo,
  Phys. Rev. Lett. {\bf 76} (1996) 3263, Nucl. Phys. B {\bf 483}
  (1997) 67.
\bibitem{Voitkiv04} A. Voitkiv, J. Phys. B {\bf 33} (2000) 1299,
  Physics Reports {\bf 392} (2004) 191, and submitted (2005).
\end{thebibliography}

\begin{thebibliography}{99}
\bibitem{JLQCD}
   JLQCD Collaboration, Phys.\ Rev.\ {\bf D66} (2002) 077501.
\bibitem{lat03}
   BGR Collaboration,\ arXiv:hep-lat/0309075.
\bibitem{CPPACS}
   CP-PACS Collaboration, Phys.\ Rev.\ {\bf D67} (2003) 014502.
\bibitem{full}
        CP-PACS Collaboration,\ Phys.Rev. D70 (2004) 074513.
\bibitem{Gattringer}
        BGR Collaboration,\ arXive:hep-lat/0409064.
\bibitem{Colangelo}
   G.~Colangelo {\it et al.}, Nucl.\ Phys.\ {\bf B603} (2001) 125.
\end{thebibliography}

\begin{thebibliography}{99}
\bibitem{R98.01}    PSI experiment R--98.01, http://pihydrogen.web.psi.ch.
\bibitem{Lyu00}     V.~E.~Lyubovitzkij and A.~Rusetski, Phys. Lett. B 494 (2000) 9.
\bibitem{Gas03}     J.~Gasser et al., Eur. Phys. J. C 26 (2003) 13.
\bibitem{Zemp04}    P.~Zemp, this workshop.
\bibitem{Eri02}     T.~E.~O.~Ericson et al., Phys. Rev. C 66 (2002) 014005; this workshop.
\bibitem{Hen03}     M.~Hennebach, thesis Universit\"at zu K\"oln, 2003.
\bibitem{Hadatom03} Proceedings of HadAtom 2003, http://www.hadatom05.unibe.ch/.
\bibitem{meson04}   D.~Gotta et al., Int. J. Mod. Phys. vol 20 (2005) 349.
\bibitem{Sch01}     H.--Ch.~Schr\"oder et al., Eur. Phys. J. C 21 (2001) 433.
\bibitem{ECRIT}     D.~F.~Anagnostopoulos et al., Nucl. Instr. Meth. B 205, 9 (2003).
\bibitem{Jensen05}  T.~Jensen, this workshop.
\end{thebibliography}

\begin{thebibliography}{12}

\bibitem{PSI}
D.~Chatellard {\it et al.},
Nucl.\ Phys.\ A {\bf 625} (1997) 855;
P.~Hauser {\it et al.},
Phys.\ Rev.\ C {\bf 58} (1998) 1869.

\bibitem{Meissner}
S.~R.~Beane, V.~Bernard, T.-S.~H.~Lee and U.-G.~Mei{\ss}ner,
Phys.\ Rev.\ C {\bf 57} (1998) 424
[arXiv:nucl-th/9708035].
S.~R.~Beane, V.~Bernard, E.~Epelbaum, U.-G.~Mei{\ss}ner and D.~R.~Phillips,
Nucl.\ Phys.\ A {\bf 720} (2003) 399
[arXiv:hep-ph/0206219].



\bibitem{Borasoy}
B.~Borasoy and H.~W.~Grie{\ss}hammer,
Int.\ J.\ Mod.\ Phys.\ E {\bf 12} (2003) 65 [arXiv:nucl-th/0105048].

\bibitem{Beane}
S.~R.~Beane and M.~J.~Savage,
Nucl.\ Phys.\ A {\bf 717} (2003) 104
[arXiv:nucl-th/0204046].

\bibitem{deuteron}
U.-G. Mei{\ss}ner, U. Raha and A. Rusetsky,
arXiv:nucl-th/0501073.

\end{thebibliography}

\begin{thebibliography}{99}

\bibitem{Arndt}
R.A.~Arndt, W.J.~Briscoe, I.I.~Strakovsky and R.L.~Workman,
Phys. Rev. {\bf C69} (2004) 035213.

\bibitem{Bugg}
D.V.~Bugg,
Eur. Phys. J. {\bf C33} (2004) 505.

\bibitem{Ericson}
T.E.O.~Ericson, B.~Loiseau and S.~Wycech,
Phys. Lett. {\bf B594} (2004) 76.

\bibitem{MES}
M.E.~Sainio, Eur. Phys. J. {\bf A24} (2005), s2, 89.

\end{thebibliography}

\begin{thebibliography}{99}

\bibitem{potential}
  S. Deser, M. L. Goldberger, K. Baumann and W. Thirring,
  {\it Phys.\ Rev.\ }{\bf 96} (1954) 774;
  %
  G. Rasche and W. S. Woolcock,
  {\it Helv.\ Phys.\ Acta.\ }{\bf 49} (1976) 557;
  %
  D.~Sigg {\it et al.},
  {\it Phys.\ Rev.\ Lett.\ }{\bf 75} (1995) 3245.

\vspace{-.2cm}

\bibitem{Lyubovitskij}
  V.~E.~Lyubovitskij and A.~Rusetsky,
  {\it Phys.\ Lett.\ }{\bf B 494} (2000) 9 [arXiv:hep-ph/0009206];
  %
  J.~Gasser, M.~A.~Ivanov, E.~Lipartia, M.~Mojzis and A.~Rusetsky,
  {\it Eur.\ Phys.\ J.\ }{\bf C 26} (2002) 13  [arXiv:hep-ph/0206068].

\vspace{-.2cm}

\bibitem{Zemp}
  P.~Zemp, "Pionic Hydrogen in QCD $\boldsymbol{+}$ QED: Decay width at NNLO",
  {\it PhD Thesis, University of Bern} (2004); 
  %
  J.~Gasser and P.~Zemp, to be published.

\vspace{-.2cm}

\bibitem{Eiras}
  D.~Eiras and J.~Soto,
  {\it Phys.\ Lett.\ }{\bf B 491} (2000) 101,
  [arXiv:hep-ph/0005066].

\vspace{-.2cm}

\bibitem{Buettiker}
  P.~B\"uttiker {\it et al.}, work in progress.

\vspace{-.2cm}

\bibitem{Ericson}
  T.~E.~O.~Ericson, B.~Loiseau and S.~Wycech,
  {\it Phys.\ Lett.\ }{\bf B 594} (2004) 76;
  %
  %
  A.~N.~Ivanov, M.~Faber, A.~Hirtl, J.~Marton and N.~I.~Troitskaya,
  {\it Eur.\ Phys.\ J.\ }{\bf A 18} (2003) 653
  [arXiv:nucl-th/0306047].

\end{thebibliography}

\vspace{-.20in}
\begin{thebibliography}{99}
\bibitem{ERI04}
T. E. O. Ericson,
B. Loiseau, and S. Wycech,  
Phys. Lett. B {\bf 594} (2004) 76.
\bibitem{ERI05} T. E. O. Ericson and A. Ivanov, to appear.
\bibitem{JG02}
J. Gasser, {\it et al.},
 Eur. Phys. J. C{\bf 26}, 13 (2002).
\bibitem{TE88} 
T. E. O. Ericson and W. Weise, 
in {\it Pions and Nuclei}, Clarendon Press, Oxford, 1988.
\end{thebibliography}

\begin{thebibliography}{99}
\bibitem{psiexp}
  D. Gotta, contribution to this workshop.  
\bibitem{bracci78}
  L.~Bracci and G.~Fiorentini, 
  Nuovo Cim. {\bf 43A} (1978) 9.
\bibitem{kilic04}
  S.~Kilic, J.-P. Karr, and L. Hilico, 
  Phys. Rev. A {\bf 70} (2004)  042506.
\bibitem{jensen02}
  T.S. Jensen and V.E.~Markushin,
  Eur. Phys. J. D {\bf 21} (2002) 271.
\bibitem{jensen03} 
  T.S. Jensen and V.E. Markushin, 
  in  \textit{Precision Physics of Simple Atomic Systems},
  Eds. S.G.~Karshenboim; V.B.~Smirnov,
  Lecture Notes in Physics {\bf 627} (Springer, Berlin 2003) 37.
\bibitem{jensen04} 
  T.S. Jensen, Eur. Phys. J. D {\bf 31} (2004) 11. 

\end{thebibliography}

\begin{thebibliography}{99}

\bibitem{pNRQED} A. Pineda and J. Soto, Nucl. Phys. (Proc. Suppl.) {\bf 64}, 428
  (1998). 

\bibitem{HBET}
  J.~Gasser and H.~Leutwyler,
  Annals Phys.\  {\bf 158}, 142 (1984); 
  E. Jenkins and A.V. Manohar, Phys. Lett. {\bf B255}, 558
  (1991). 


\bibitem{Pineda:2002as}
  A.~Pineda,
  Phys.\ Rev.\ C {\bf 67}, 025201 (2003); hep-ph/0308193; hep-ph/0412142.
  
\bibitem{Taqqu} F. Kottmann et al., Hyperfine Interact. {\bf 138}, 55 (2001). 


\end{thebibliography}

\begin{thebibliography}{99}

\bibitem{Ishiwatari04}T. Ishiwatari et al., Phys. Lett. B593
(2004) 48.
\bibitem{pubH} G. Beer et al., submitted for publication.
\bibitem{iwa}M. Iwasaki et al., Phys. Rev. Lett. 78 (1997) 3067.

\end{thebibliography}

\begin{thebibliography}{99}

\bibitem{Baldini} 
R. Baldini et al. [DEAR Collaboration], DAPHNE exotic atom research: The DEAR
proposal, LNF-95-055-IR.

\bibitem{Cargnelli}
M. Cargnelli et al. [DEAR Collaboration], in Proceedings of the HadAtom03
Workshop, 12-17 October 2003,ECT(Trento, Italy),
[arXiv:hep-ph/0401204].

\bibitem{Gall}
A. Gall, J. Gasser, V. E. Lyubovitskij, A. Rusetsky,
Phys. Lett. B {\bf 462} (1999) 335
[arXiv:hep-ph/9905309].

\bibitem{Lyubovitskij}
V. E. Lyubovitskij, A. Rusetsky,
Phys. Lett. B {\bf 494} (2000) 9
[arXiv:hep-ph/0009206].

\bibitem{Zemp}
P. Zemp, in Proceedings of HadAtom03 Workshop, 13-17 October 2003 ECT(Trento
Italy), J. Gasser, P. Zemp, in preparation,
[arXiv:hep-ph/0009206].

\bibitem{Schweizer}
J. Schweizer,
Eur. Phys. J. C {\bf 36} (2004) 483
[arXiv:hep-ph/0405034].

\bibitem{Meissner}
U.-G. Mei\ss ner, U. Raha, and A. Rusetsky
Eur. Phys. J. C {\bf 35} (2004) 349
[arXiv:hep-ph/0402261].

\end{thebibliography}

\begin{thebibliography}{12}
\bibitem{IV4}
A. N. Ivanov {\it et al.},
Eur. Phys. J. A {\bf 23},  79  (2005), nucl--th/0406053.
\bibitem{TE88} 
T. E. O. Ericson and  W. Weise, 
in {\it Pions and Nuclei}, Clarendon Press, Oxford, 1988.
\end{thebibliography}

\begin{thebibliography}{99}
\bibitem{Gasser:1987rb} 
J.~Gasser, M.~E.~Sainio and A.~Svarc,
Nucl. Phys. B {\bf 307} (1988) 779.
\bibitem{Becher:1999he}
T.~Becher and H.~Leutwyler,
Eur.\ Phys.\ J.\ C {\bf 9} (1999) 643 
[arXiv:hep-ph/9901384]; 
JHEP {\bf 0106} (2001) 017 
[arXiv:hep-ph/0103263].
\bibitem{Kubis:2000zd}
B.~Kubis and U.~G.~Meissner,
Nucl.\ Phys.\ A {\bf 679} (2001) 698
[arXiv:hep-ph/0007056]. 
\bibitem{PCQM} 
V.~E.~Lyubovitskij, T.~Gutsche and A.~Faessler,
Phys.\ Rev.\ C {\bf 64} (2001) 065203 [arXiv: hep-ph/0105043];   
A.~Faessler, T.~Gutsche, V.~E.~Lyubovitskij, K.~Pumsa-ard, 
Prog. \ Part. \ Nucl. \ Phys. {\bf 54} (2005) (in print).  
\bibitem{Beg:1973sc}
M.~A.~B.~Beg and A.~Zepeda,
Phys.\ Rev.\ D {\bf 6} (1972) 2912.
\end{thebibliography}

\begin{thebibliography}{99}

\bibitem{CI}
N.~Cabibbo and G.~Isidori,
  arXiv:hep-ph/0502130.

\bibitem{Silagadze}
 Z.~K.~Silagadze,
  JETP Lett.\  {\bf 60} (1994) 689
  [arXiv:hep-ph/9411382].

\end{thebibliography}

\begin{thebibliography}{99}

\bibitem{Moliere} G. Moli\'ere, Z.Naturforsh, {\bf 3a} (1948) 78

\bibitem{Fano}  U.Fano,  Phys.Rev. {\bf 93}(1954) p.117

\end{thebibliography}

\begin{thebibliography}{99}

\bibitem{Nem:HA03} L. Nemenov, talk at HadAtom03, arXiv:hep-ph/0401204

\bibitem{vosk:bert} C. Bertulani, G. Baur, Electromagnetic processes
  in relativistic heavy ion collisions, Physics Report vol. 163
  (1988) 299--408

\end{thebibliography}

\begin{thebibliography}{99}

\bibitem{Fri02}E. Friedman, Phys. Lett. B {\bf 524} (2002) 87; 
Nucl. Phys. A {\bf 710} (2002) 117.

\bibitem{Wei01}W. Weise, Nucl. Phys. A {\bf 690} (2001) 98c.

\bibitem{ETa82}T.E.O. Ericson and L. Tauscher, Phys. Lett. B {\bf 112}
(1982)425.

\bibitem{FGa04}E. Friedman and A. Gal, Phys. Lett. B {\bf 578} (2004) 85.

\bibitem{FBB04}E. Friedman et. al., 
Phys. Rev. Lett. {\bf 93} (2004) 122302.

\end{thebibliography}

\begin{thebibliography}{99}

\bibitem{} All the speakers of the workshop have contributed to one or 
several of the topics mentioned in the table - 
I do not list them individually.

\end{thebibliography}
